\documentclass[aps,prl,twocolumn,superscriptaddress]{revtex4}

\RequirePackage{doi}
\usepackage{hyperref}
\usepackage{graphicx}
\usepackage{bm}
\usepackage{amsmath}
\usepackage{amssymb}
\usepackage{hyperref}

\begin{document}
\title{Low-temperature friction on suspended graphene: Negative friction?}
	
\author{Zhao Wang}
\email{zw@gxu.edu.cn}
\affiliation{Department of Physics, Guangxi University, 530004 Nanning, China}

\begin{abstract}
Molecular dynamics is used to simulate the sliding of a diamond-like-carbon tip on a suspended graphene layer. The average force acting on the tip from the graphene along the sliding direction is measured to be negative at liquid-helium temperature. This ``negative'' friction is attributed to a spontaneous transverse oscillation of the suspended graphene according to a corrugation of the potential surface. The oscillation induces a transverse force which contributes importantly to the total energy dissipation at 4K. These results suggest that, in nanotribology experiments, the friction force should not be considered only to present in the sliding direction, even though there is no net relative displacement along the transverse/lateral direction between the two surfaces in contact.
\end{abstract}
		
\maketitle

Friction is a critical problem for nanodevices with moving components. This is due to a dramatic increase in the surface-to-volume ratio when the component size lies in the nanoscale, a domain in which macroscale lubrication methods have been deemed inadequate \cite{Kendall1994}. Recent advances in the development of nanoelectromechanical systems (NEMS) have spurred renewed efforts to search for new lubricants that can work at the nanoscale \cite{Kim2007}. Graphene, a monoatomic layer of carbon, is believed to be a particularly good candidate for such an ideal lubricant thanks to its chemical inertness, extreme mechanical strength and peculiar two-dimensional (2D) structure. Distinct frictional properties have been reported for graphene in different tribosystems \cite{Vanossi2013,Wu2021}. In particular, Kawai \textit{et al.} demonstrated the superlubricity of graphene nanoribbons on gold at liquid-helium temperature \cite{Kawai2016}. Intensive experimental and theoretical efforts were made to understand the lubricating behaviors of graphene on different types of substrates \cite{Egberts2014,Klemenz2014,Li2016,Meng2020,Wang2019c,Wang2018,Wang2019d}. It was shown that thermally-induced structural distortion to the out-of-plane direction at medium or high temperature is large enough to dramatically suppress the superlubricity. However, the tribological properties of graphene at low temperature remain poorly understood. 

Single-asperity contact measurements have been a useful tool in nanotribology \cite{Szlufarska2008}. In such experiments, a sharp tip with a radius typically between 10 and 100 nm is controlled to slide on the sample's surface for measuring friction force. Using a nanoscale tip sliding on a chemically modified graphite surface, Deng \textit{et al.} reported a negative coefficient of friction in the low-load regime, which is attributed to a reversible partial delamination of the
topmost atomic layers \cite{Deng2012,Thormann2013}. Similar effects were also reported by a number of related recent works \cite{Mandelli2019a,Kwang-Hua2020,Liu2020}. The reported negative friction coefficient is based on a
negative slope of the curve of friction force vs load, which can be explained by a model where the adhesion
between the atomic force microscope (AFM) tip and the substrate surface is stronger than the dispersion forces holding together the graphite layers. However, the friction force can never be negative, as obliged by the law of energy conservation. 

In most single-asperity experiments, the friction is measured in the sliding direction of the tip. The force in the transverse direction is commonly assumed to be negligible, because it is hard to measure. The present molecular dynamics (MD) simulations however demonstrate that this assumption becomes problematic at extremely low temperature. By simulating sliding of a diamond-like-carbon (DLC) tip atop of a suspended graphene layer, we show that the force acting on the tip in the sliding direction can even be negative at liquid-helium temperature.


\begin{figure}[htp]
\centerline{\includegraphics[width=9cm]{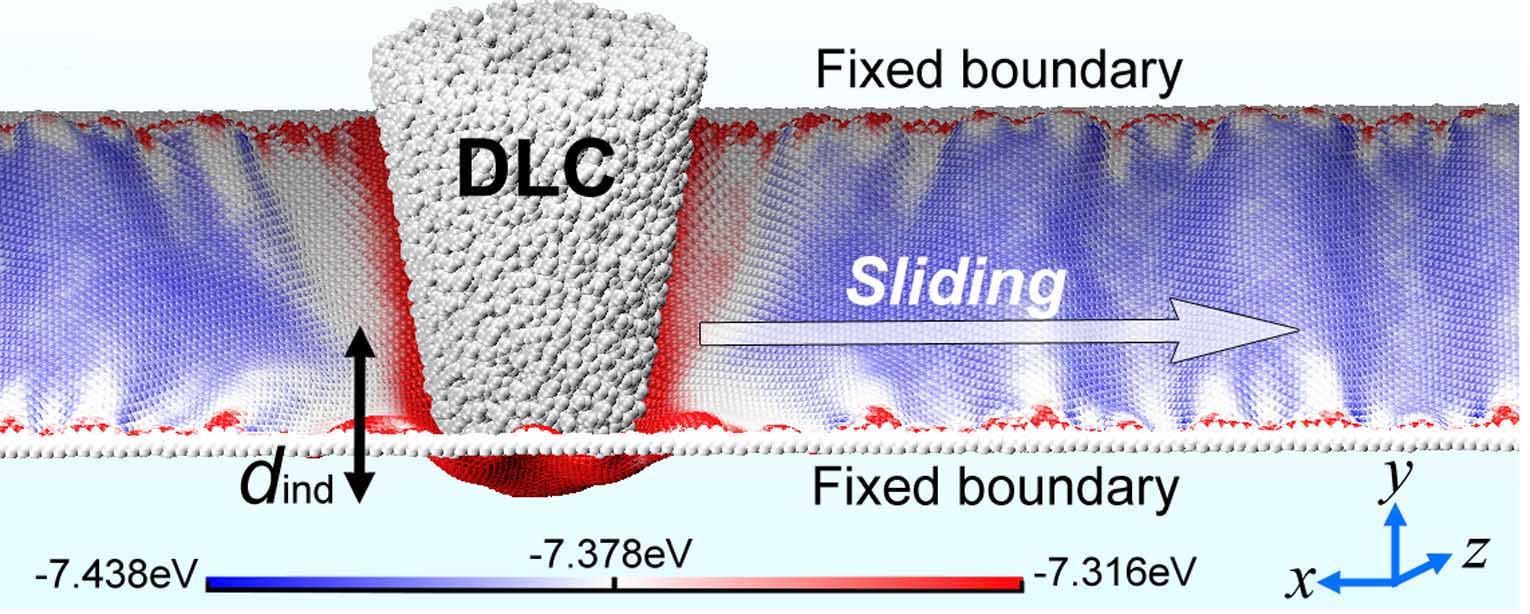}}
\caption{\label{F1}
Model setup for a DLC tip sliding on a graphene layer suspended between two parallel hypothetical supports with a trench width of $20\;\mathrm{nm}$). Color bar corresponds to the atomic potential energy.}
\end{figure}

A suspended graphene layer constitutes an ideal sample since the tribological properties of graphene have been proven to be sensitive to the presence of supports \cite{Li2016}. In our MD simulations, a DLC tip indents on the central axis of a graphene layer that is suspended between two parallel supports, and then slides along the the trench, as shown in Fig.\ref{F1}(a). We here define three orthogonal axis as $x$ in the sliding direction, $y$ in the lateral direction, and $z$ in the transverse direction.

The atomistic configuration of the DLC tip with a spherical-bottom geometry ($4$nm in diameter) is generated by melting a single-crystal nanodiamond at $5000\;\mathrm{K}$ followed by a rapid quenching process with a cooling rate of $5 \times 10^{5}\;\mathrm{K/ns}$ \cite{Sha2013}. The tip slides at a constant speed of $0.04\;\mathrm{nm/ps}$, while being kept rigid since the out-of-plane stiffness of the graphene is much lower than that of the DLC \cite{Bosak2007}. The Nos\'{e}-Hoover thermostat is used to equilibrate the system at different temperatures (liquid-helium $\sim 4.125\;\mathrm{K}$, liquid-nitrogen $\sim 77\;\mathrm{K}$, $\sim 300\;\mathrm{K}$ and $\sim 600\;\mathrm{K}$).

The simulation is performed with the parallel molecular dynamics package LAMMPS \cite{Plimpton95}. The interactions between the carbon atoms are mimicked in the framework of the adaptive interatomic reactive empirical bond order (AIREBO) potential \cite{Stuart2000a}. The total interatomic potential involves many-body terms as a collection of that of individual bonds. The long-range interactions are included by adding a parameterized Lennard-Jones 12-6 potential term with a cutoff radius of $1.0\;\mathrm{nm}$. This potential has intensively been used in the literature for many simulation works on the mechanical properties of carbon nano-materials. It involves a many-body bond-torsion term and therefore enables a smooth transition from long-range interaction to chemical bonding. It hence affords a good description to the structural flexibility and bond formation (or breaking) in carbon nanostructures \cite{Wang2019a,Qi2018,Wang2009d,Wang2009c}.


\begin{figure}[htp]
\centerline{\includegraphics[width=9cm]{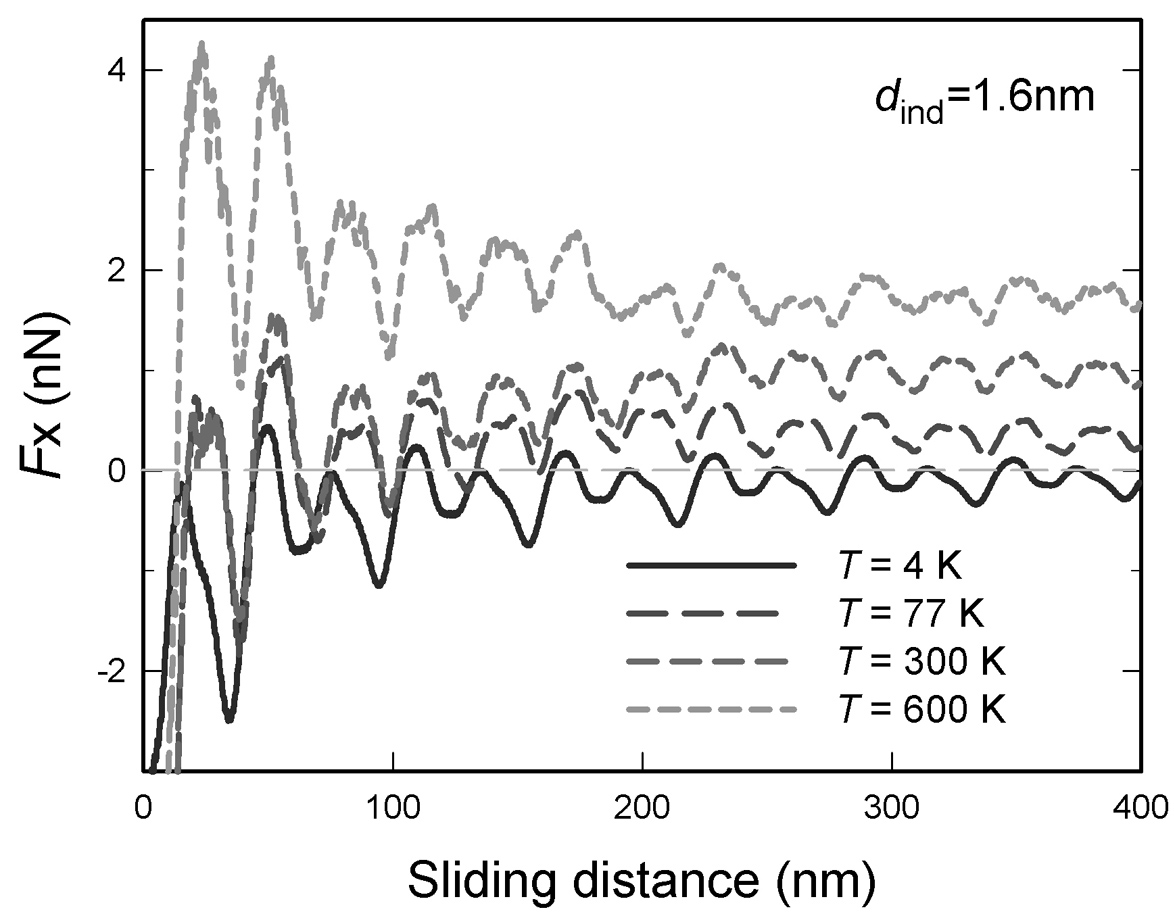}}
\caption{\label{F2}
Time-averaged force from the graphene sensed by the tip along $x$ vs sliding distance for a constant indentation depth $d_{ind}=1.6 \;\mathrm{nm}$. Each line-style represents a case at a different temperature.}
\end{figure}

Fig.\ref{F2} shows the variation of the force acting on the tip in the sliding direction $F_{x}$ as a function of the sliding distance at different temperatures. It is seen that the oscillation amplitude of $F_{x}$ is large at the beginning of the simulations and it then tends to stabilize, a behavior is consistent with the experimental observation of Zhang \textit{et al.} \cite{Zhang2015a}, with magnitudes comparable to previous measurements  \cite{Deng2013,Vilhena2016,Long2017}. We see that $F_{x}$ decreases with decreasing temperature, and strikingly, it becomes negative at $4\;\mathrm{K}$. This is different from the previously-reported negative friction coefficient due to a curve with a negative slope when the friction force is plotted against load \cite{Deng2012}, and cannot be explained by the state-of-art theory of adhesive friction \cite{Gao2004}. 
   
\begin{figure}[htp]
\centerline{\includegraphics[width=9cm]{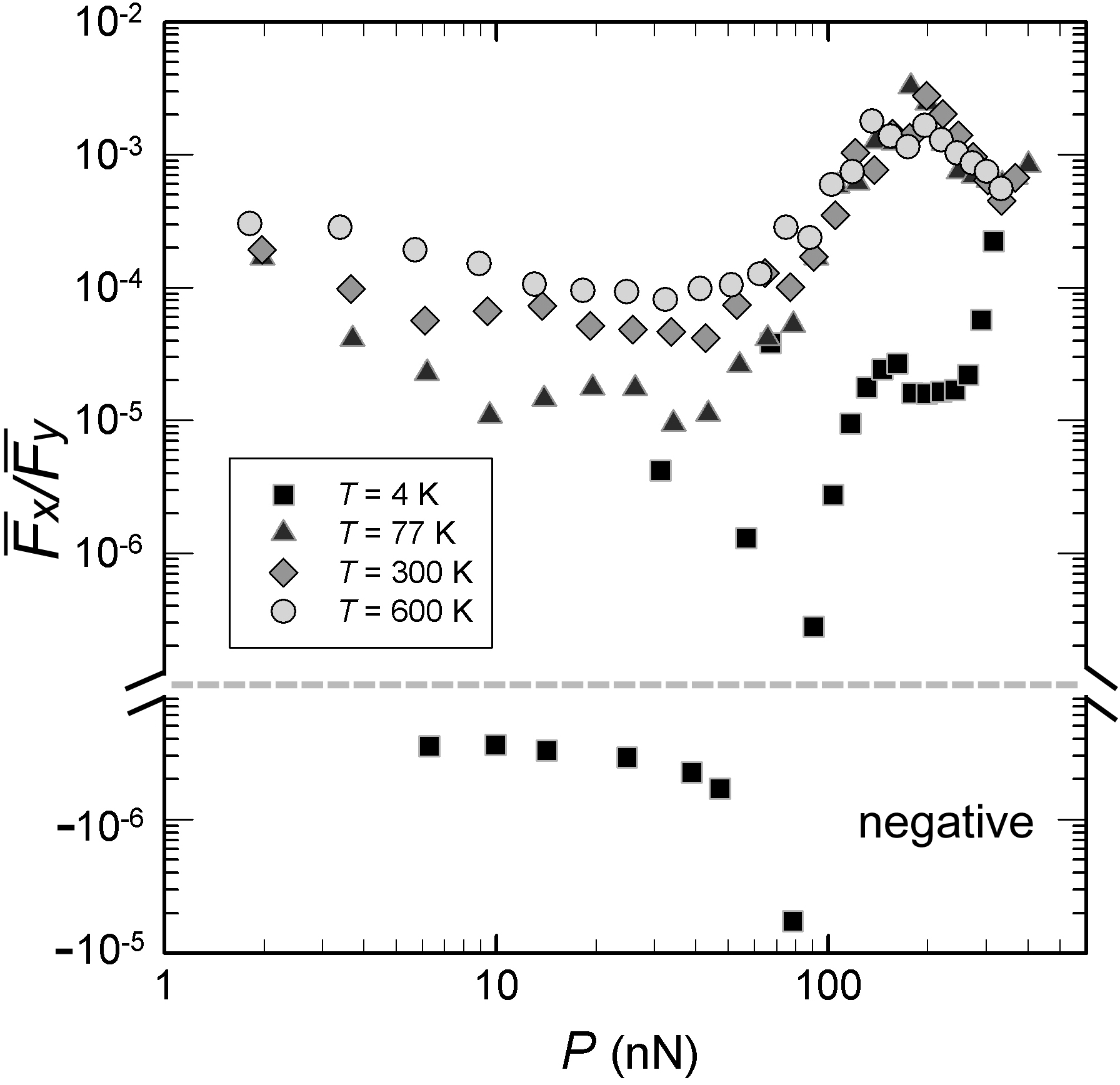}}
\caption{\label{F3}
$\bar{F}_{x}/\bar{F}_{y}$ vs normal load $\bar{F}_{y}$ at different temperatures.}
\end{figure}	
	
The forces acting on the tip from the graphene are time-averaged to measure the mean force $\bar{F}_{x}$ and the normal load $\bar{F}_{y}$. Their ratio is plotted in Fig.\ref{F3} in relation to $P$. It is seen that $\bar{F}_{x}/\bar{F}_{y}$ decreases and then increases at raising temperatures and tends to saturate beyond a high load. As a particular feature, $\bar{F}_{x}/\bar{F}_{y}$ at $4\;\mathrm{K}$ are found to be negative for $\bar{F}_{y}<90 \;\mathrm{nN}$. Beyond this threshold, $\bar{F}_{x}/\bar{F}_{y}$ becomes positive and rapidly increases with the raising load before saturating at a value about $1.6 \cdot 10^{-5}$ for $\bar{F}_{y} > 180 \;\mathrm{nN}$ before the rupture occurs. The value of $\bar{F}_{x}/\bar{F}_{y}$ seems to be a reference to the kinetic coefficient of friction, however, at low temperature this is proven not to be the case since $\bar{F}_{x}$ is negative and it cannot be the true friction force.

\begin{figure}[htp]
\centerline{\includegraphics[width=9cm]{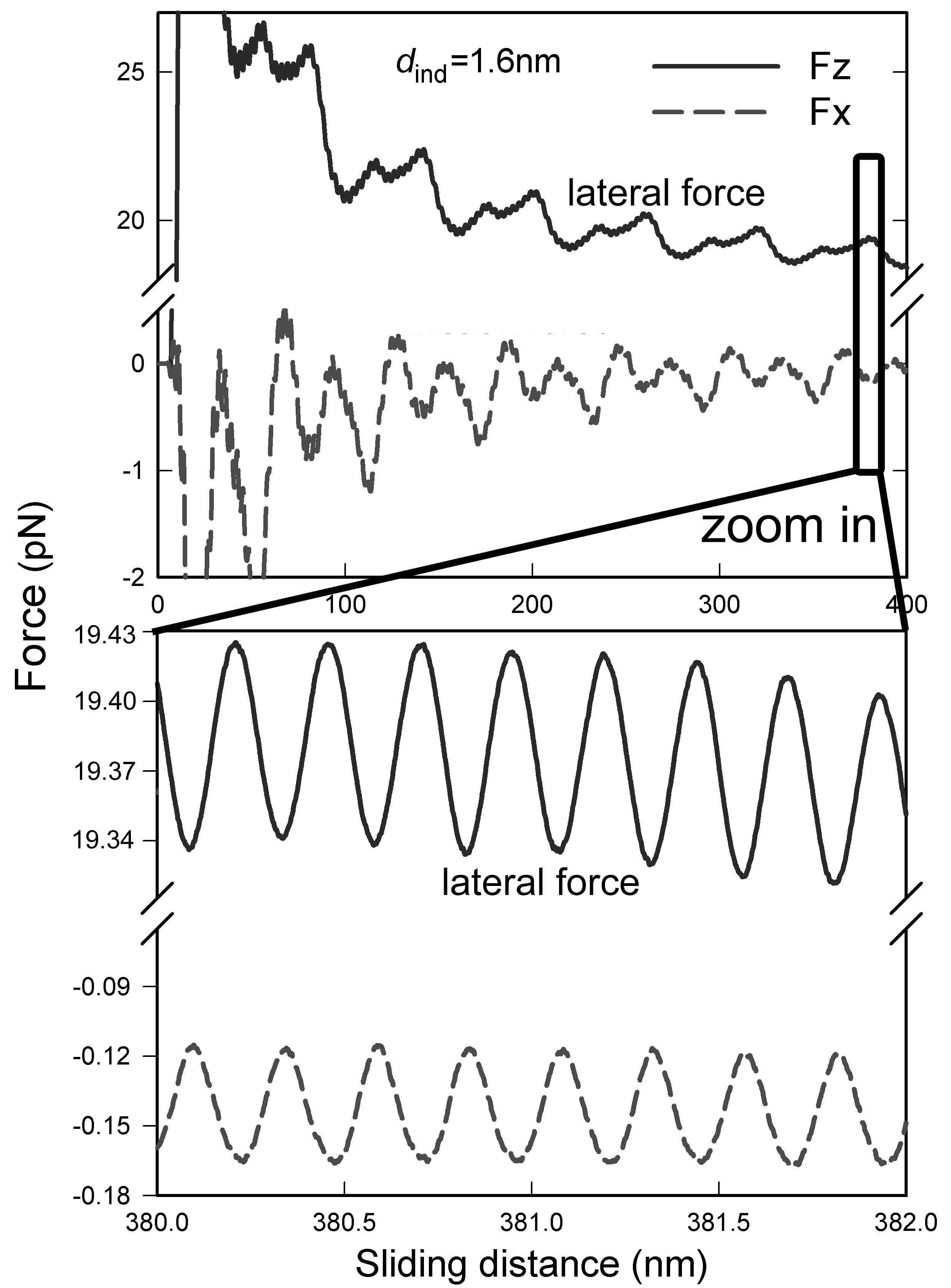}}
\caption{\label{F4}
Instantaneous forces along the sliding $(x)$ and transverse $(z)$ direction acting on the tip as a function of the sliding distance at $4\;\mathrm{K}$ for an indentation depth $d_{ind}=1.6 \;\mathrm{nm}$. The lower panel is a zoom-in view of the upper panel.}
\end{figure}

\begin{figure}[htp]
\centerline{\includegraphics[width=7.8cm]{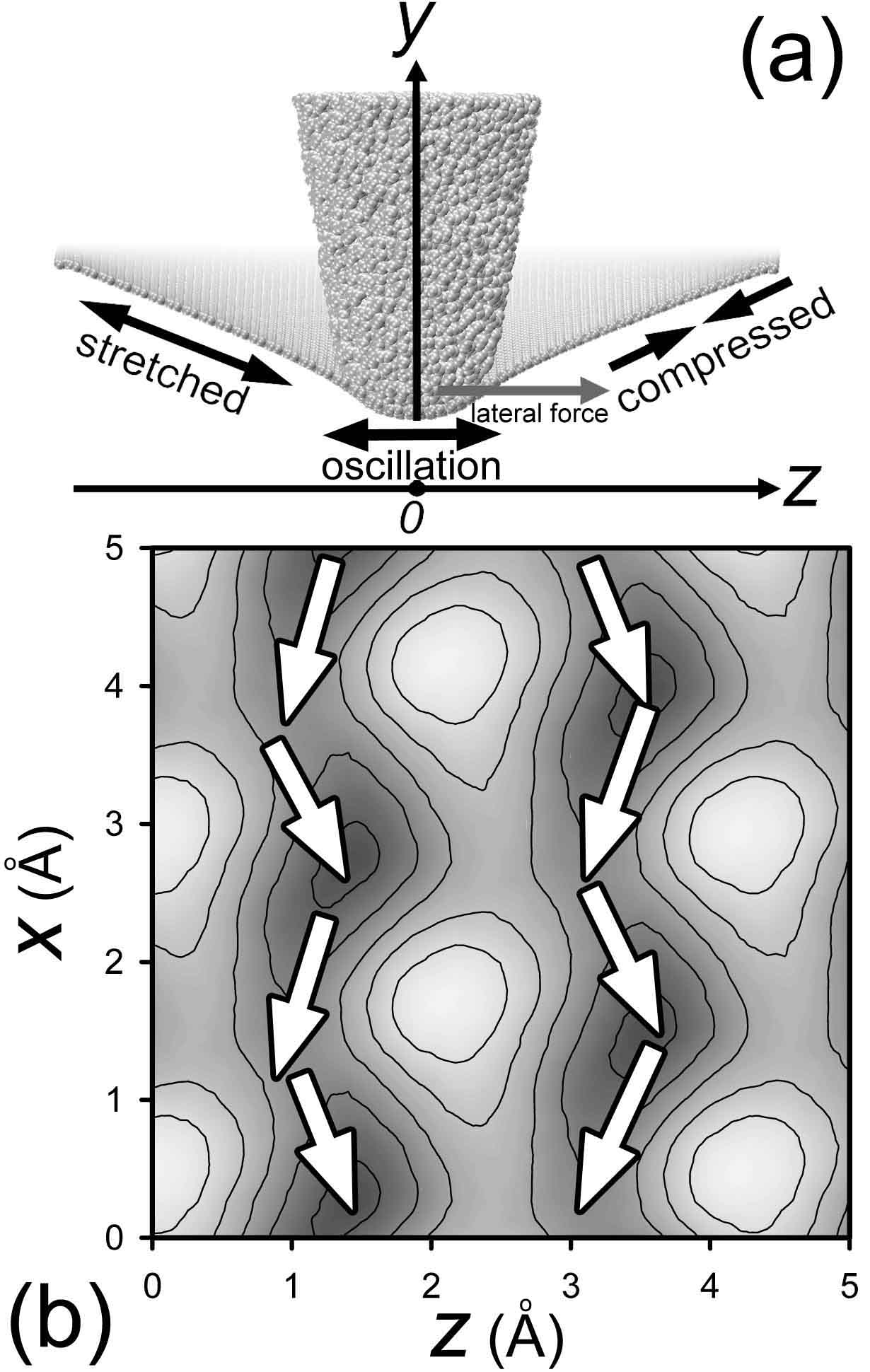}}
\caption{\label{F5}
(a) Schematic for the transverse force induced perpendicular to the sliding direction. (b) Distribution of the potential energy for the DLC-graphene interaction. The arrows point on an energetically-favorable sliding trajectory of the tip. The potential profile is acquired by displacing the DLC tip atop the graphene in discretized steps of $0.02\;$\AA \cite{Verhoeven2004}.}
\end{figure}

So, what could be the true friction force? Does a negative $\bar{F}_{x}$ breaks the energy conservation law? Toward  answers to these questions, we plot in Fig.\ref{F4} instantaneous forces along the sliding $(x)$ and the transverse $(z)$ directions acting on the tip as a function of the sliding distance at $4\;\mathrm{K}$ for an indentation depth $d_{ind}=1.6 \;\mathrm{nm}$. It is seen that the variation of the transverse force $F_{z}$ is at about the same order of magnitude of that of $F_{x}$. Apparently, the alternating $F_{z}$ is the key to the negative $F_{x}$, since the friction should be calculated in the direction of relative displacement between the two contacting bodies. $F_{z}$ is associated with an oscillation of the suspended graphene with respect to the DLC tip [Fig.\ref{F5} (a)], which is induced by the asymmetric DLC-graphene interaction potential landscape as shown in Fig.\ref{F5} (b). 

It can be seen that the energy landscape is asymmetric with respect to the highest-energy point due to the amorphous structure of the DLC tip. This leads to a non-conventional stick-slip behavior, as it is energetically favorable for the tip to slide along a sinusoidal path in the ``valleys'' of the potential profile for a minimum energy corrugation, as shown by the arrows in the Fig.\ref{F5} (b). Since the tip is rigid, the suspended graphene has to oscillate for following the sinusoidal path. i.e. the graphene atoms are energetically obliged to displace in the direction normal to the sliding direction along not only $y$ but also $z$ axis, in order to accommodate tip movements along such a sinusoidal trajectory of the tip. This displacement breaks the geometrical symmetry of the graphene layer with respect to its central axis ($z=0$), and induce a net transverse force $F_{z}$ acting on the tip. Specifically, $F_{z}$ is the missing component to the friction force due to the transverse displacement of the substrate. 

A number of previous studies showed that the superlubricity of graphene originates from the incommensurate configuration of the contact, and is thus highly sensitivity to the change in the graphene surface geometry, in particular to the rippling induced by thermal fluctuation \cite{Fasolino2007,Wang2011a}. The distribution of the potential energy on a surface will be blurred by thermal fluctuation when the temperature rises, so the tip does not need to exactly follow the optimized sinusoidal paths.

By definition, friction should be measured in the direction of the relative motion of the two surfaces in contact. In the example demonstrated above, there is no net relative motion of the two surfaces in the transverse ($z$) direction. However, an oscillation occurs in $z$ following a potential corrugation. This oscillation is ultrafast (in G-THz) with tiny magnitude (in Angstrom), and is hence very difficult to be observed by mechanical means in experiments. It should universally present for all solid contacts at any temperature, since all interfaces have potential corrugation. Fortunately, this phenomenon appears only to manifest itself at extremely-low temperature, by resulting in a striking negative force in the sliding direction that is routinely assumed to be the friction in most experiments. The present work shows however that this force actually misses a hidden transverse component for being the true friction. 

\bibliographystyle{habbrv}

\end{document}